\documentclass[prb,twocolumn,showpacs,nofootinbib]{revtex4}
\usepackage[dvips]{graphicx}
\usepackage{color,array,dcolumn}
\usepackage{amsmath}
\usepackage{amssymb}

\begin{document}

\title{Magnetization of a homogeneous two dimensional fermion gas with repulsive contact interaction and Rashba spin-orbit potential}
\author{A. Ambrosetti}
\email{alberto.ambrosetti@unipd.it}
\author{P.L. Silvestrelli}
\email{pierluigi.silvestrelli@unipd.it}
\author{F. Toigo}
\email{flavio.toigo@unipd.it}
\affiliation{Dipartimento di Fisica, University of Padova, via Marzolo 8, I--35131, 
Padova, Italy}

\begin{abstract}
\date{\today}
The $z$ axis magnetization of a two dimensional electron gas with contact repulsive interaction
and in presence of a Rashba potential is computed by means of quantum field theory at
second order. A striking effect of the pure Rashba interaction is that of hindering spin 
alignment along the $z$ direction. Evidence of transition at critical repulsive interaction
coupling constant is still found. The degree of magnetization, however, shows a clear dependence
on the spin-orbit interaction strength. Furthermore, the transition to magnetized state appears to
be smoothed by the presence of the Rashba interaction.
\end{abstract}

\maketitle

\section{Introduction}
The magnetization of a homogeneous gas of fermions
is a long standing problem, already considered by several authors during the past years
\cite{ceperley,attaccalite,pilati,macdonald,
cond3d}. It is well known, for instance, how, in presence of a contact repulsive
interaction, both the two and three dimensional spin $1/2$ fermion gas undergo a phase
transition, magnetizing upon increase of the repulsive potential coupling constant \cite{macdonald,conduitprl,belitz}. A first, tentative, experimental proof of itinerant ferromagnetism in cold fermion atoms was
recently given by Jo $et$ $al.$ \cite{exp1,exp2}
The basic mechanism underlying the phase transition could be well understood in terms of a
competition between kinetic energy and repulsion of fermions with opposite spins. 
For strong repulsive interactions the potential energy reduction induced by
the polarization will compensate the corresponding increase of kinetic energy, favoring
the appearance of a magnetization.
\newline
Though the physical behavior of the above mentioned systems appears to be well understood,
the recent interest in spin orbit (SO) interactions both in semiconductor devices and in
systems of cold atoms, leads to new questions regarding the appearance of 
possible unknown spin effects.
Among the different SO interactions present in semiconductors, the Rashba interaction \cite{Rashba}
certainly plays a fundamental role, especially due to its tunable strength.
The form of the Rashba interaction is:
\begin{equation}
V_{Rashba}=\lambda(k_y\sigma_x-k_x\sigma_y).  \\
\end{equation}
where $k_{x,y}$ represents the momentum operator and $\sigma_{x,y}$ are Pauli matrices.
This SO interaction generates in asymmetric quantum wells due to the existence
of an electric field perpendicular to the plane of confinement \cite{Kouwenhoven,Steward,nitta,engels,espab}. The confinement introduced by the
very narrow quantum well, in turn, allows for a two dimensional treatment of the electrons,
which are described as effectively moving in the $x-y$ plane.
Moreover, an external gate potential could be applied in order to effectively modify the strength
of the interaction $\lambda$ \cite{nittathick}.
\newline
Recently, the Rashba interaction has been reproduced in ultracold atoms \cite{Dalibard, cold} by means of 
controlled laser beams externally applied to the system. The realization of SO interactions by means
of artificial gauge fields certainly boosted the interest on SO interactions, and a number of
theoretical studies appeared in the last few months, regarding SO effects in superfluid or
superconducting states \cite{salas,guo,iskin,agarwal}.
Experimentally, a quasi two dimensional cold atom gas can be realized by means of 
counterpropagating laser beams along the $z$ direction with antinodes at half wavelength
spacing $b$.
The ultracold atom gas is a particularly favorable system for the study of
magnetization properties. The absence of interfering phonons, present in solid state systems,
and the elevated degree of control achievable, make  it a valuable tool for investigating
delicate magnetic properties, such as combined SO and population imbalance effects.
\newline
Some of the most relevant questions related to the two dimensional fermion gas in presence of
Rashba interaction are those related to its magnetization properties. The interplay of Rashba
interaction and magnetization appears indeed not yet well understood. A recent Diffusion Monte
Carlo simulation\cite{io} and perturbative analytical approaches\cite{pert} for the two 
dimensional electron gas in presence of both Coulomb repulsion and
Rashba interaction revealed negligible two-body effects on the occupation of single particle
Rashba states and no appreciable magnetization along the $z$ axis.
In the following we will consider an unpolarized cold atoms assembly  with a repulsive two body interaction. 
Since due to the very low-energy only s-wave scattering is important, the interaction may be modeled by a contact interaction acting  only  between particles of opposite spins, due to Pauli principle.  Therefore, in the action of the uniform 2-D system confined to the $x-y$  plane it will be described by a term of the form:

\begin{equation}
V=g \sum_{\mathbf{k}_1,\mathbf{k}_2,\mathbf{q}}
\bar{\psi}_{\mathbf{k}_1+
\mathbf{q}}^{\uparrow}\bar{\psi}_{\mathbf{k}_2-\mathbf{q}}^{\downarrow}\psi_{\mathbf{k}_2}^{\downarrow}\psi_{\mathbf{k}_2}^{\uparrow}
\label{interaction}
\end{equation}

The  coupling constant $g$ is assumed to be positive,
and $\psi^{\uparrow , \downarrow}$ indicate the fermion fields with $\pm 1/2$ $z$ spin component.
Experimentally,  $g$ may be modified by exploiting the
Feshbach resonance mechanism. \cite{feshbach}
Notice that the interaction in Eq. ({\ref {interaction}) differs from the Coulomb one between electrons in a quantum well, as it  lacks  the long range 
tail and acts only between particles with opposite spin.
\newline
The present paper is organized as follows:
In the II section, the path integral computation of the free energy at second order in the
coupling constant $g$ will be illustrated. In particular, results will be derived considering
the possible presence of an external Zeeman potential, introducing a chemical potential
difference, i.e. an unbalance, between the $\uparrow$ and $\downarrow$ populations.
Some analytical results will be discussed regarding the case of zero external potential.
In Section III numerical results will be shown for the magnetization of the system as a
function of both the $\lambda$ and the $g$ coupling constants.

\section{Second order perturbation theory}
In the following, a path integral derivation of the system free energy will be given, up
to second order in the repulsive potential coupling constant $g$, while exactly including the
Rashba effects in the independent particle propagator.
We stress that the second order approximation to the free energy will be obtained as an expansion
in the two body interaction, starting from the independent particle solutions.
This approach is to be preferred in two dimensions with respect to a saddle point approximation.
In fact, in absence of Rashba interaction, the 
stationary points of the action do not correspond to its minima but instead they provide maxima and therefore one cannot 
proceed to evaluate the partition function by considering only small fluctuations around them, since large fluctuations would be favored instead.
A minimum of the energy, however, is correctly recovered by a minimization of the energy calculated by standard perturbation theory, starting from the solution
of the non interacting system.
This may be  understood by expressing the energy at first order in $g$ in terms of the spin $\uparrow$ and spin
$\downarrow$ densities:
\begin{equation}
E/V=\frac{\pi}{m}(n_{\uparrow}^2+n_{\downarrow}^2)+gn_{\uparrow}n_{\downarrow}.
\end{equation}
In the 2D gas, both the kinetic energy and the repulsive term show quadratic dependence
 As a consequence, for a given at $n= n_{\uparrow}+n_{\downarrow}$  it will be a parabolic function of the net magnetization $s=n_{\uparrow}-n_{\downarrow}$, with a  minimum or a maximum at $s=$ according to whether $\frac{\pi}{m}-g$ is positive or negative, respectively. So the stationary value at $s=0$ is a maximum for large values of $g$ while the energy is obtained when either $n_{\uparrow}=n$ or $n_{\downarrow}=n$, i.e. when the system is completely polarized.
The independent particle propagator ${\mathcal{G}}_0$ in presence of Rashba SO, could be obtained by inverting the following
expression for ${\mathcal{G}}_0^{-1}$, written in the $\uparrow, \downarrow$ spin basis:
\begin{eqnarray}
{\mathcal{G}}_0^{-1}(\omega_n,\mathbf{k})= \qquad \qquad \qquad \qquad \qquad \qquad \qquad  \qquad \qquad \\ \nonumber 
 = \left(
  \begin{array}{cc}
        -i\omega_n+\frac{k^2}{2m}-\mu-\Delta\mu     &  \lambda(k_y+ik_x)  \\
	 \lambda(k_y-ik_x)             & -i\omega_n+\frac{k^2}{2m}-\mu+\Delta\mu  
  \end{array} \right)
\end{eqnarray}
where $\omega_n=\frac{\pi}{\beta}(2n+1)$ are fermions Matsubara frequencies,  $\mu=\frac{\mu_{\uparrow}+\mu_{\downarrow}}{2}$ is the chemical potential of the system, while 
$ \Delta \mu=\frac{\mu_{\uparrow}-\mu_{\downarrow}}{2}$.
An external potential of the form $-\Delta \mu \sigma_z$ is included,
favoring the occupation of $\uparrow$ states with respect to $\downarrow$.
The diagonalization of ${\mathcal{G}}_0^{-1}$ leads to the eigenenergies
\begin{equation}
\epsilon_{\pm}=\frac{k^2}{2m}\pm\sqrt{\Delta\mu^2+\lambda^2k^2}
\end{equation}
corresponding to the eigenstates
\begin{equation}
\phi^{\pm}(\mathbf{k})= c_{\pm,\mathbf{k}}
  \left(
    \begin{array}{c}
    \frac{\lambda(k_y+ik_x)}{\Delta \mu\pm \sqrt{\Delta \mu^2+\lambda^2k^2}}     \\
    1
  \end{array} \right)
\label{rstates}
\end{equation}
where the normalization constants $c_{\pm,\mathbf{k}}$ are defined to be real and obey the equation:
\begin{equation}
c_{\pm,\mathbf{k}}^2\left( \frac{\lambda^2 k^2}{(\Delta \mu\pm \sqrt{\Delta \mu^2+\lambda^2k^2})^2} +1 \right)=1.
\end{equation}
These states will be hereafter referred to as $\pm$  for simplicity. 
Knowing the ${\mathcal{G}}_0^{-1}$ eigenstates, it is possible to write the transformation matrix
\begin{equation}
\mathcal{R}_{\mathbf{k}}=
  \left(
  \begin{array}{cc}
        c_{+,\mathbf{k}}\frac{\lambda(k_y-ik_x)}{\Delta \mu + \sqrt{\Delta \mu^2+\lambda^2k^2}}     &   c_{+,\mathbf{k}}  \\
        c_{-,\mathbf{k}}\frac{\lambda(k_y-ik_x)}{\Delta \mu - \sqrt{\Delta \mu^2+\lambda^2k^2}}     &   c_{-,\mathbf{k}}  \\
  \end{array} \right)
\end{equation}
which diagonalizes the independent particle inverse propagator, yielding
\begin{equation}
{\mathcal{G}}^{-1}_{0,diag}(\omega,\mathbf{k})=\mathcal{R}_{\mathbf{k}}{\mathcal{G}}_0^{-1}(\omega,\mathbf{k})\mathcal{R}^{\dagger}_{\mathbf{k}}
\label{gdiag}
\end{equation}
where $\mathcal{G}^{-1}_{0,diag}$ is the diagonalized independent particle inverse.
We stress that the inclusion of the Rashba interaction in the independent particle propagator
allows for a non perturbative treatment of the SO interaction. 
\newline
A perturbative approach is instead employed for the two body repulsive interaction.
By using the relation
\begin{equation}
\bar{\psi}_{\uparrow}\bar{\psi}_{\downarrow}\psi_{\downarrow}\psi_{\uparrow}=
\frac{1}{4}\left(\sum_{\alpha=\uparrow,\downarrow} \bar{\psi}_{\alpha} \psi_{\alpha}\right)^2
-\frac{1}{4}\left(\sum_{\alpha,\beta=\uparrow,\downarrow} \bar{\psi}_{\alpha}\sigma^z_{\alpha\beta} \psi_{\beta}\right)^2
\label{inter}
\end{equation}
it is possible to employ a double Hubbard-Stratonovich transformation after introducing the
auxiliary fields $\rho$ and $\phi_z$,  associated to density and $z$ magnetization respectively:
\begin{eqnarray}
e^{-\frac{g}{4}(\bar{\psi}_{\alpha}\delta_{\alpha\beta}\psi_{\beta})^2}=
\int d\rho e^{g\rho^2-g\rho \bar{\psi}_{\alpha}\delta_{\alpha\beta}\psi_{\beta}} \nonumber \\
e^{\frac{g}{4}(\bar{\psi}_{\alpha}\sigma^z_{\alpha\beta}\psi_{\beta})^2}=
\int d\phi^z e^{-g(\phi_z)^2+g\phi_z \bar{\psi}_{\alpha}\sigma^z_{\alpha\beta}\psi_{\beta}}
\label{strat}
\end{eqnarray}
The grand canonical partition function can then be expressed as
\begin{equation}
Z=\int D\phi_z D\rho D\bar{\psi} D\psi
e^{-\int d\tau d\mathbf{r}g(\phi_z^2-\rho^2)
+{\bar{\psi_{\alpha}}}[{\mathcal {G}}^{-1}_{0\alpha\beta}+g\delta_{\alpha\beta}-g\sigma^z_{\alpha\beta}\phi_z]\psi_{\beta}}.
\label{part}
\end{equation}
Integrating over the fermion fields and expanding the action up to quadratic order in the
fields $\phi_z$ and $\rho$ one obtains
\begin{eqnarray}
&&S(\phi_z,\rho)=\int d\tau d\mathbf{r} \{ \ g(\phi_z^2-\rho^2)-Tr ln {\mathcal{G}}_0^{-1}- \nonumber \\ 
&& -g Tr[{\mathcal{G}}_0(\rho {\bf 1}-\phi_z \sigma^z)]+\nonumber \\
&&+\frac{g^2}{2}Tr[{\mathcal{G}}_0(\rho {\bf 1}-\phi_z \sigma^z){\mathcal{G}}_0(\rho {\bf1}-\phi_z \sigma^z)]+O(g^3) \}. \qquad
\label{act}
\end{eqnarray}

Then, by expressing ${\mathcal{G}}_0$ in terms of $\mathcal{G}_{0 diag}$ and $\mathcal{R}$ (see \eqref{gdiag}),
it is possible to rewrite the second line of the above equation as:
\begin{equation}
\mathbf{L}\cdot\mathbf{\Phi}(k=0)
\end{equation}
where $\mathbf{\Phi}=(\rho,\phi_z)$ and
\begin{equation}
\mathbf{L}^T=
  g\beta \sum_{\mathbf{k}}\left(
  \begin{array}{c}
        f(\xi^+_k)+f(\xi^-_k)  \\
        f(\xi^+_{\mathbf{k}})(1-2c_{+,\mathbf{k}}^2)+f(\xi^-_{\mathbf{k}} )(1-2c_{-,\mathbf{k}}^2)  \\
  \end{array} \right)
\end{equation}
Following a similar, though slightly more complex procedure, the third line of \eqref{act}
is expressed as:
\begin{equation}
\sum_{\bf q}\mathbf{\Phi}^T_{-{\bf q}} \mathcal{Q}_{\bf q}\mathbf{\Phi}_{\bf q}
\end{equation}
with:
\begin{equation}
\mathcal{Q}_{\bf q}=\frac{1}{2}Tr_{\bf k}
  \left(
  \begin{array}{cc}
        \mathcal{M}     &  \mathcal{N}  \\
	\mathcal{N}     &  \mathcal{P}  
  \end{array} \right) \,
\label{cu}
\end{equation}
where the trace $Tr_{\bf k}$ is taken over two fermion momenta ${\bf k}_1$ and ${\bf k}_2$ satisfying the
relation ${\bf k}_2-{\bf k}_1=-{\bf q}$. $\mathcal{M}$ stands for 
\begin{equation}
\mathcal{M}_{k_1,k_2}=\sum_{s_1,s_2= \pm }m_{s_1,s_2,k_1,k_2}{\mathcal G}^{0s_1}_{k_1}{\mathcal G}^{0s_2}_{k_2}
\end{equation}
with similar expressions for $\mathcal{N}$ and $\mathcal{P}$.
${\mathcal{G}}_0^+$ and ${\mathcal{G}}_0^-$ are the + and - components of ${\mathcal G}_{0diag}$ and $n$, $m$ and $p$ are
defined as follows:
\begin{eqnarray}
n_{s_1,s_2,k_1,k_2}=\frac{(c_{s_1,\bf{k_1}}\lambda k_1)^2}{(\Delta\mu+s_1\sqrt{\Delta\mu^2+\lambda^2k_1^2})^2}\cdot \nonumber \qquad \qquad \\ 
\cdot \frac{(c_{s_2,\bf{k_2}}\lambda k_2)^2}{(\Delta\mu+s_2\sqrt{\Delta\mu^2+\lambda^2k_2^2})^2} -(c_{s_1,\bf{k_1}}c_{s_2,\bf{k_2}})^2 \nonumber \\
m_{s_1,s_2,k_1,k_2}=n_{s_1,s_2,k_1,k_2}+d_{1,s_1,s_2,k_1,k_2}+d_{2,s_1,s_2,k_1,k_2}\nonumber \\
p_{s_1,s_2,k_1,k_2}=n_{s_1,s_2,k_1,k_2}+d_{1,s_1,s_2,k_1,k_2}-d_{2,s_1,s_2,k_1,k_2}\nonumber \\
\end{eqnarray}
with
\begin{eqnarray}
d_{1,s_1,s_2,k_1,k_2}=2(c_{s_1,\bf{k_1}}c_{s_2,\bf{k_2}})^2\nonumber \qquad \qquad \qquad \qquad\\ 
d_{2,s_1,s_2,k_1,k_2}=2\frac{(c_{s_1,\bf{k_1}}c_{s_2,\bf{k_2}}\lambda)^2}{(\Delta\mu+s_1\sqrt{\Delta\mu^2+\lambda^2k_1^2})}\cdot \nonumber \\
\cdot \frac{{\bf k}_1 \cdot {\bf k}_2 }{(\Delta\mu+s_2\sqrt{\Delta\mu^2+\lambda^2k_2^2})} \qquad \qquad
\end{eqnarray}
The last term on the first line of \eqref{act} could also be accounted for
by modifying the matrix $\mathcal{Q}$ into
\begin{equation}
{\mathcal Q}'_q=Tr_{\bf k}
  \left(
  \begin{array}{cc}
        -g\beta V {\bf 1} +\frac{g^2}{2}\mathcal{M}     &  \frac{g^2}{2}\mathcal{N}  \\
	\frac{g^2}{2}\mathcal{N}     & g\beta V {\bf 1} +\frac{g^2}{2}\mathcal{P}  
  \end{array} \right)
\end{equation}
The partition function \eqref{part} can thus be written as
\begin{eqnarray}
Z=\int D\phi^z D\rho \sim e^{-S_0} \int D\phi^z D\rho e^{-\mathbf{\Phi}^T {\mathcal Q}' \mathbf{\Phi}-\mathbf{L}^T\mathbf{\Phi}}= \nonumber \\
=e^{-S_0}(Det {\mathcal Q}')^{-1/2}e^{\frac{1}{4}\mathbf{L}^T {\mathcal Q}'^{-1}\mathbf{L}} \qquad
\label{partition}
\end{eqnarray}
where $S_0$ represents the action of the independent particle system. 
The logarithm of the determinant of $\mathcal{Q}'$ will then be expanded in $g$ and
only terms up to quadratic order will be retained.
The above expression contains a multiplicity of terms and certainly appears
more complicated than the corresponding formula, obtained in absence of
Rashba interaction.
The reason of this complication resides in the momentum dependent transformations
${\mathcal{R}}_{\bf k}$, which needs to be taken into account due to the spin structure of the
Rashba independent particle solutions.

\section{Polarization}
The polarization along the $z$ axis  defined as
\begin{equation}
M=\frac{n_{\uparrow}-n_{\downarrow}}{n} ,
\label{magnetization}
\end{equation}
where 
 $n_{\uparrow , \downarrow}$ are the  density components with spin  $\uparrow$ or $\downarrow$ and $n$ is the total density, 
is simply   $2\langle\phi^z \rangle /n$  with:
\begin{eqnarray}
\langle \phi^z \rangle=\frac{\left( \int D\phi^z D\rho e^{-S+J\phi^z}\phi^z\right)_{J\rightarrow 0}}{\left( \int D\phi^z D\rho e^{-S+J\phi^z}\right)_{Ji\rightarrow 0}}= \nonumber \\
=\frac{\partial}{\partial J} ln \left( \int D\phi^z D\rho e^{-S+J\phi^z}\phi^z\right)_{J\rightarrow 0}
\end{eqnarray}
The introduction of the field $J$ corresponds to modifying the vector
$\mathbf{L}$ into 
\begin{equation}
\mathbf{L}'=\mathbf{L}+
\left(
 \begin{array}{c}
        0  \\
	J  
  \end{array} \right)
\label{elle}
\end{equation}
which then yields
\begin{eqnarray}
\langle \phi^z \rangle = \frac{1}{4\beta V} \big[ \left(0,1\right) {\mathcal Q}'^{-1}\mathbf{L}+\mathbf{L}^T {\mathcal Q}'^{-1}
\left(
 \begin{array}{c}
        0  \\
	1  
  \end{array} \right)
\big]= \nonumber \\
\frac{1}{2\beta^2V^3g}\Big[ \left(\frac{g^2}{2}Tr \mathcal{N} \right) \left( \sum_k f(\xi^+_{\mathbf{k}})+f(\xi^-_{\mathbf{k}})\right)+ \nonumber \\
+\left(g\beta V-\frac{g^2}{2}Tr  \mathcal{M} \right)\cdot \nonumber \\
\cdot \left( \sum_{\mathbf{k}} f(\xi^+_{\mathbf{k}})(1-2c_{+,\mathbf{k}}^2)+f(\xi^-_{\mathbf{k}} )(1-2c_{-,\mathbf{k}}^2)\right)
\Big]
\end{eqnarray}
\noindent From the above formula it follows that if $\Delta\mu=0$ then  $c_{\pm , \mathbf{k}}=1/\sqrt{2}$ and the $z$ polarization, is identically zero.

We stress that this result is not sufficient in order to establish the absence of magnetization
for any value of the coupling constant $g$, since it is a direct consequence of the fact that the expectation value of $ \langle \sigma_z \rangle $ on both our $+$ and $-$ unperturbed  Rashba states (see Eq. (\ref {rstates})) is zero, when $\Delta\mu=0$. Therefore even when the occupation numbers of these states are different, the total magnetization  is zero.
To allow a nonzero magnetization, we must start from unperturbed Rashba states where $ \langle \sigma_z \rangle $  may be nonzero.
This is easily achieved by introducing a variational field favoring the occupation of $\uparrow$ over $\downarrow$ spin states, i.e. substituting
$\Delta \mu$ with $\Delta \mu+h $, in all the above formulas.
A the end of the calculation one will get $h$ by minimizing  $E-h M n$ .

As will be discussed in the subsequent section, in general one will find that when $\Delta \mu=0$, the minimum energy is achieved for 
$h=0$ only when the repulsion strength $g$ is below a critical value $g_c$.  No magnetization will thus be present
when $g < g_c$ .
\newline
As mentioned above, this property appears to be closely related to the single particle spin properties of the system: 
in fact, when $h=0$,  the expectation value of $\sigma_z$ over any of the Rashba states \eqref{rstates}
(at $\Delta\mu=0$)  is equal to zero
due to spin rotation around the particle wave vector axis \cite{lipp}.
Remarkably, in this case the independent particle Rashba states contain no dependence on the parameter
$\lambda$, also implying constant $ \langle \sigma_z \rangle=0$ even at $\lambda \rightarrow 0$.
\newline
In order to gain a more complete picture of the whole magnetization properties of the gas,
calculations were also performed for $ \langle \sigma_x \rangle $ and $ \langle \sigma_y \rangle $.
In fact, the absence of a $z$ magnetization in presence of Rashba SO coupling does not in 
principle exclude the appearance of non zero in-plane magnetization.
As an example, in the 2D fermi gas in absence of Rashba interaction, due to spin-rotation 
invariance, magnetization might equally occur along any direction if $\Delta \mu=0$ ,
while being necessarily oriented along $z$ for $\Delta \mu \neq 0$. \cite{cond2d}
\newline
In order to describe the possible occurrence of in-plane magnetization, one may 
rewrite the contact repulsive interaction \eqref{inter} as \cite{cond3d}
\begin{equation}
\frac{1}{4}\left(\sum_{\alpha=\uparrow,\downarrow} \bar{\psi}_{\alpha} \psi_{\alpha}\right)^2
-\frac{1}{4}\left(\sum_{\alpha,\beta=\uparrow,\downarrow} \bar{\psi}_{\alpha}\vec{\sigma}_{\alpha\beta} \psi_{\beta}\right)^2
\end{equation}
and consequently introduce two additional auxiliary fields $\phi_x$ and $\phi_y$ as done in
\eqref{strat}, resulting in the four dimensional analogues of ${\bf L}$ \eqref{elle} and $\mathcal{Q}$ \eqref{cu}.
Given the
complexity of the calculations, these were only performed for the case $\Delta \mu=0$ at $g<g_c$
and will be discussed only  qualitatively .
The four dimensional analog of ${\bf L}$ will again show a single non-vanishing term, corresponding
to the $\rho$ field. The new matrix will instead only have non zero off diagonal elements
corresponding to the coupling between $\phi_z$ and $\rho$ and between $\phi_x$ and $\phi_y$.
As a consequence, both the expectation values of $\phi_x$ and $\phi_y$ are identically zero
at $\Delta \mu =0$.
An analogous picture occurs in the Rashba interacting independent particle model,
where both $<\sigma_x>$ and
$<\sigma_y>$ average out to zero, due to the 2D rotational symmetry of the Fermi surface.

\section{Numerical results}
The numerical results presented in this section concern magnetization and quasipolarization properties
of the system under study at $T=0$. In the following we will express lengths in units of $n^{-1/2}$,(the 2-D particle density), and energies in units of $\frac{n \hbar^2}{ m}$
where $m$ is the mass of the particles. 
Due to the elevated computational cost related to the treatment of
some of the terms proportional to $g^2$ in the free energy, the results presented in the following
were obtained retaining only the terms of the action which are linear in $g$.
In absence of SO interaction,  the second order terms were shown to shift the transition to lower $g$
,\cite{cond2d} and to increase the order of the transition from the first to continuous.
Since our system shows a smoothing of the  transition already at first order in $g$ due to the presence of SO coupling, we expect that second order terms will only 
shift $g_c$, the critical  coupling strength for the transition, to lower values,  without changing the overall features.
\newline
As already outlined above, at $\Delta \mu=0$ no $z$ magnetization is present in the system
at repulsive couplings below  $g_c$. 
However, a phase transition, analogous to that found in absence of SO, is still present, causing
the appearance of non zero magnetization at strong repulsion.
Fig. \ref{figura1} reports the magnetization as a function of the repulsion strength for different
values of $\lambda$. While no $\lambda$ dependence is appreciable at small $g$'s, above the transition
one observes a decrease of the magnetization at saturation  by increasing $\lambda$.
Another relevant feature is the smoothing of the magnetization increase upon introduction of 
SO, the smoothing being more relevant at higher $\lambda$.
The Rashba interaction is therefore expected to effectively frustrate $M$ also at
$\Delta \mu\neq 0$. This property, in fact, is confirmed by the results reported in Figs. \ref{figura2}\ref{figura3}.
Above $g_c$ a competition will be present between the magnetizing repulsion and the 
demagnetizing Rashba interaction. Magnetization will thus increase at fixed $\lambda$ by
increasing $g$.
When $g > g_c$  the variational parameter $ h $ introduced above
becomes essential for the description of magnetization.
since the energy minimum occurs in this case for $h \neq 0$,
corresponding to partial spin alignment along the $z$ axis.
\newline
Non zero M is obtained, even below criticity, for $\Delta \mu \neq 0$, increasing with the
$\Delta \mu$, as from Fig. \ref{figura3}. This result is clearly understandable,
given the role of $\Delta \mu$ in energetically favoring the occupation of $s_z=1/2$ spin states.
Notice that the magnetization enhancement due to $\Delta \mu$ decreases as the Rashba coupling constant $\lambda$ increases.

The so-called "quasi-polarization", defined as
\begin{equation}
\xi=\frac{n_{+}-n_{-}}{n_{tot}}
\end{equation}
and corresponding to the difference between the Rashba spin + and - density fraction at $\Delta \mu=0$,
shows no difference from that of independent particles, computed by retaining only the kinetic energy terms in the system 
energy. This also appears in good agreement with QMC results.
For non zero $\Delta \mu$ values, however, also the "quasi-polarization" appears to be
affected by the presence of the repulsive interaction, showing dependences on $g$ and $\Delta \mu$
qualitatively similar to those of M.
No direct relation is observed between $M$ and $\xi$. In fact, non zero $\xi$ is found
in presence of zero $M$. At $\Delta \mu\neq 0$, however, 
non zero $M$ and $\xi$ are found, both increasing with $g$. Again, however, full quasipolarization
is consistent with partial magnetization.

\section{Conclusions}
The $z$ axis magnetization of the two dimensional fermion gas in presence of contact repulsion and
Rashba SO interaction has been computed from quantum field theory. A general expression for the
free energy at second order in $g$ is obtained. As a result, we analytically found no 
magnetization at $\Delta \mu=0$,  at values of the repulsive interaction strength below a critical value $g_c$.
Above criticity magnetization is non zero and depends on both the SO strength and $g$.
Numerical results were also given at $\Delta \mu \neq 0$, showing that the system
 develops $z$ spin polarization by increasing $\Delta \mu$. Moreover, also in this case the
repulsive interaction appears to enhance the tendency to develop $z$ magnetization.
The Rashba interaction acts by frustrating $M$, suggesting a possible
application as an effective tool for controlling the system $z$ polarization.

\section{Acknowledgements}
We acknowledge Luca Salasnich, Luca dell'Anna, Francesco Pederiva and Enrico Lipparini for
useful discussions.

\begin{figure}[ht]
\vspace{0.7cm}
\centering
\includegraphics[scale=0.3]{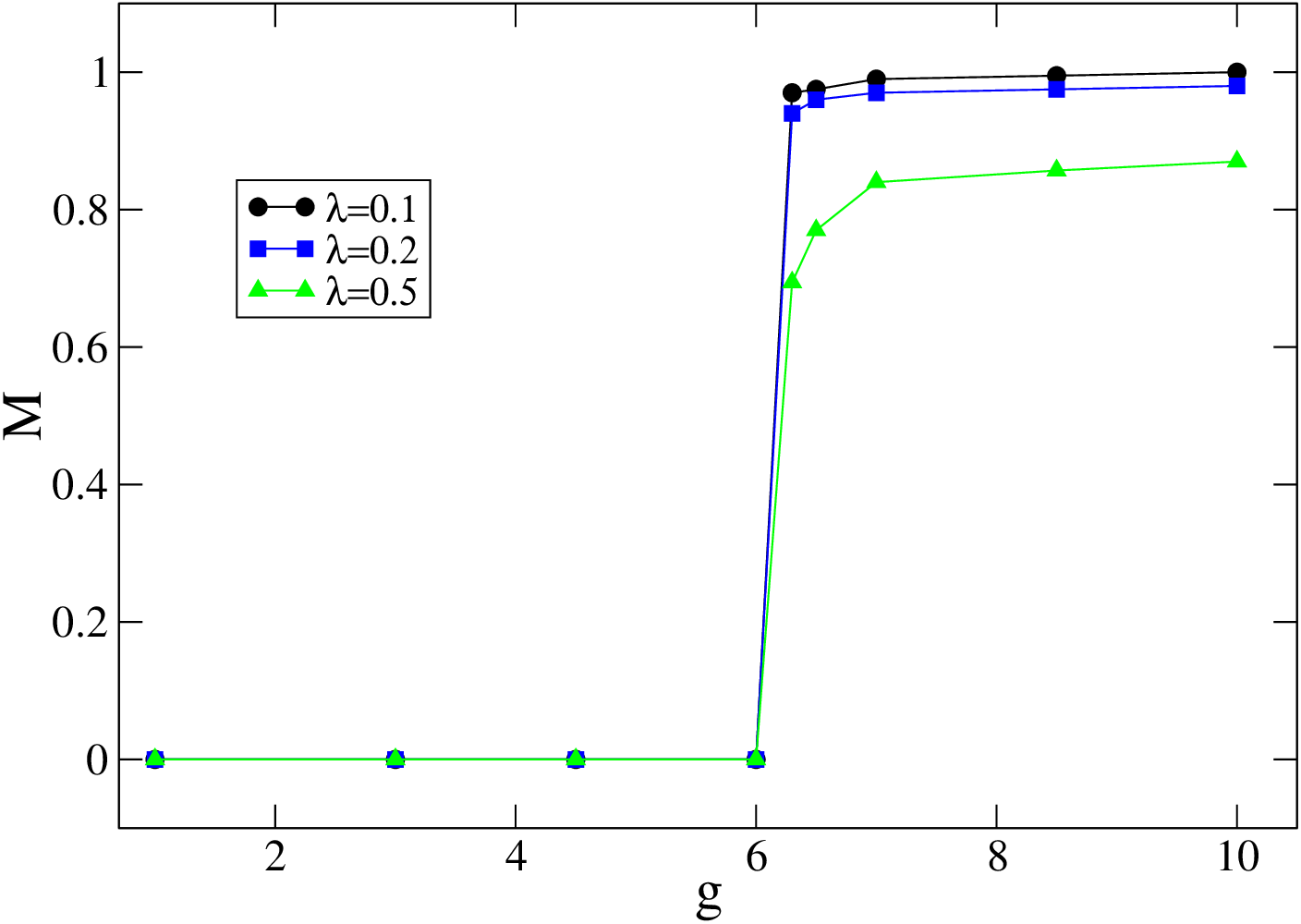}
\vspace{0.5cm}
\caption{(color online) Magnetization plotted in function of the repulsive interaction coupling constant at
fixed $\Delta \mu=0$ for different values of $\lambda$.  Units are specified in the text.}
\label{figura1}
\end{figure}

\begin{figure}[ht]
\vspace{0.7cm}
\centering
\includegraphics[scale=0.3]{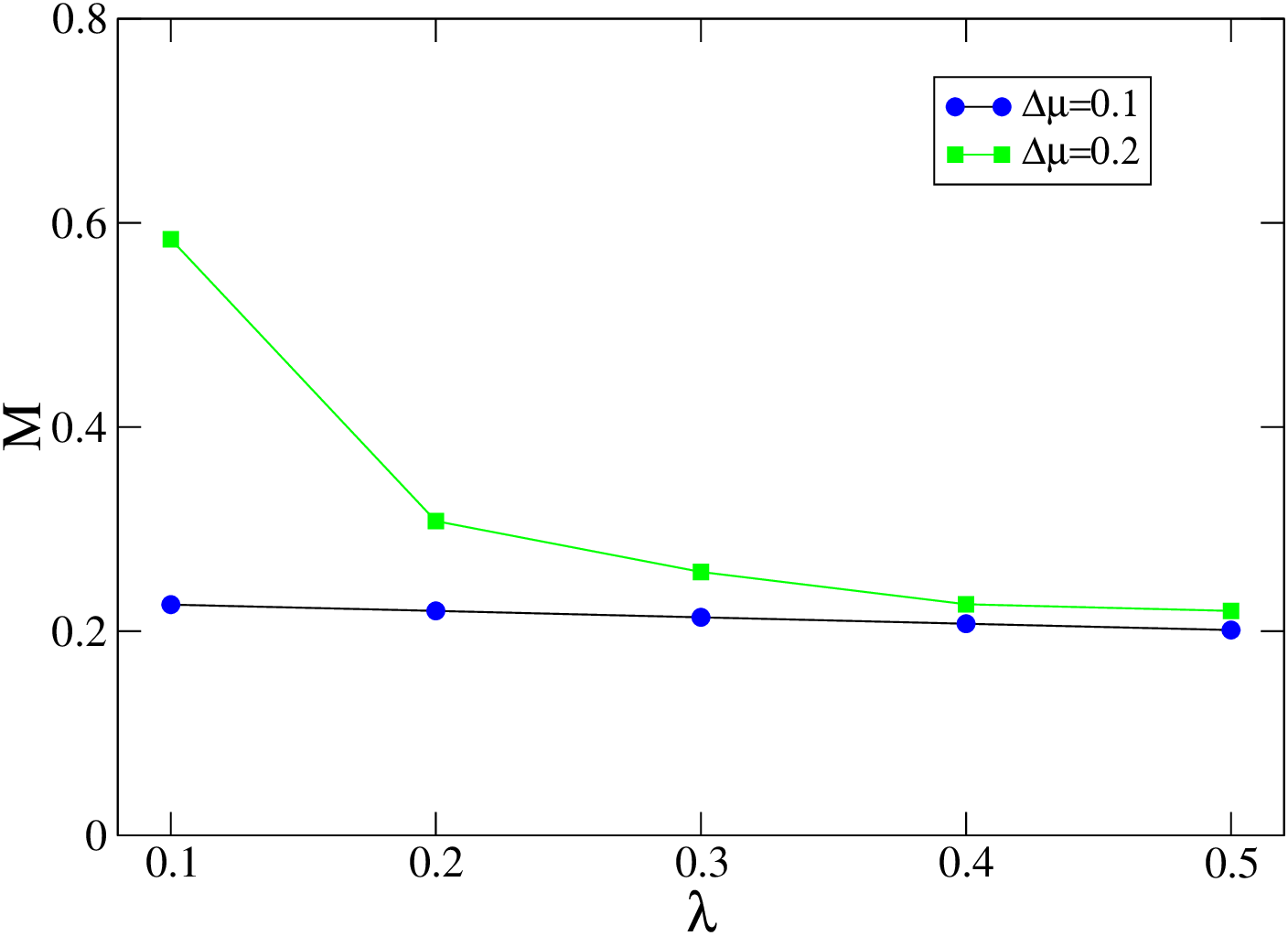}
\vspace{0.5cm}
\caption{(color online) Magnetization plotted in function of the $\lambda$ SO coupling constant
for different values of $\Delta \mu$ at fixed $g=6.0 $ Units are specified in the text}
\label{figura2}
\end{figure}

\begin{figure}[ht]
\vspace{0.7cm}
\centering
\includegraphics[scale=0.3]{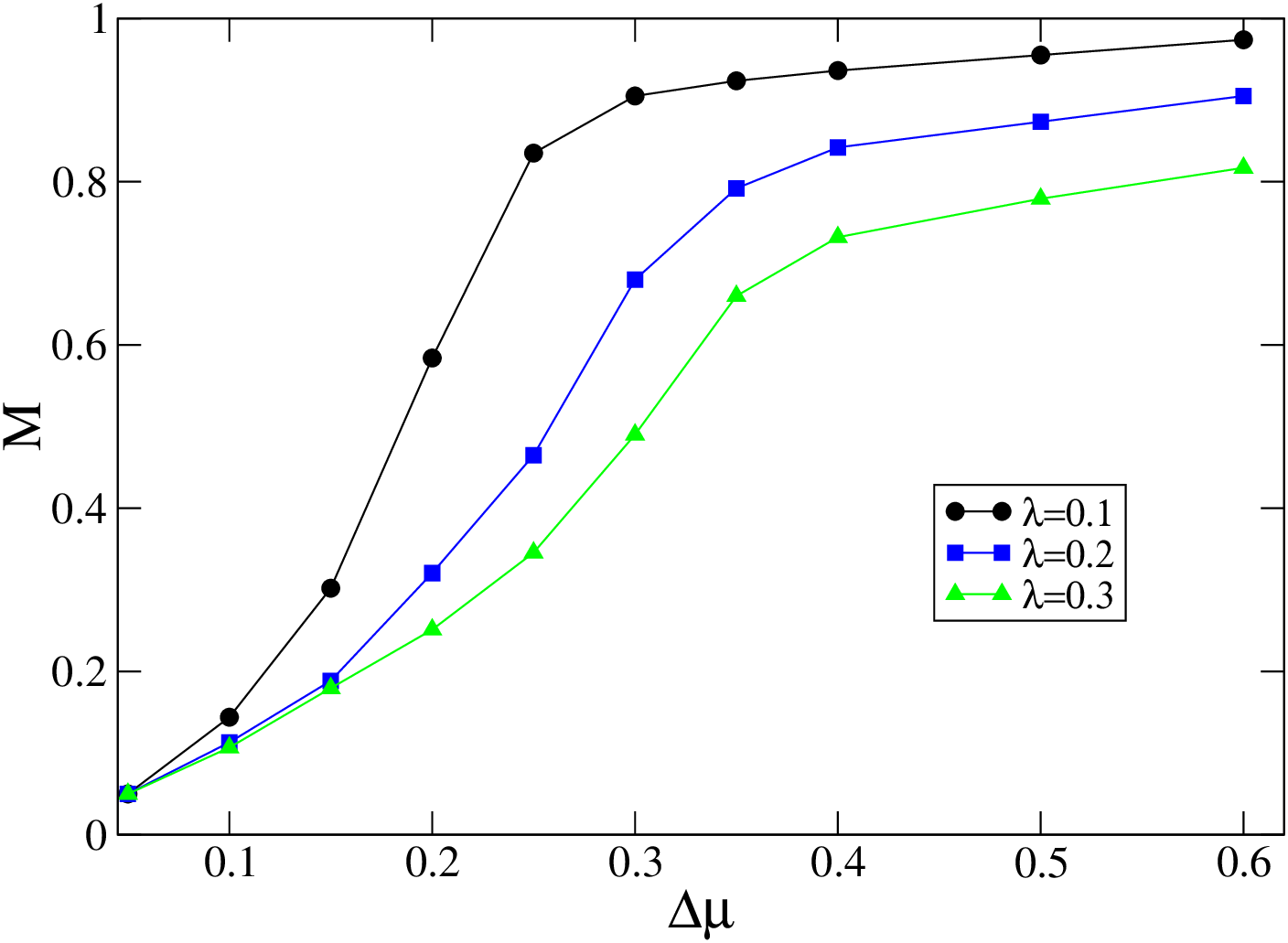}
\vspace{0.5cm}
\caption{(color online) Magnetization plotted as a function of $\Delta \mu$ at different
$\lambda$ values and fixed $g=6.0 $. Units are specified in the text.}
\label{figura3}
\end{figure}
\end{document}